# Photoferroelectric Coupling and Polarization-Controlled Interfacial Band Modulation in van der Waals Compound CuInP$_2$S$_6$

Subhashree Chatterjee, Rabindra Basnet, Rajeev Nepal, and Ramesh C. Budhani*

*Department of Physics, Morgan State University, Baltimore, MD, 21251, USA*

*ramesh.budhani@morgan.edu

**Abstract:** Understanding how optical excitation couples with polarization and interfacial electrostatics in van der Waals (vdW) ferroelectrics (FEs) is essential for the development of light-programmable nanoelectronic and optoelectronic devices. Here, we present a direct nanoscale evidence of photoferroionic coupling in the vdW FE semiconductor CuInP$_2$S$_6$ (CIPS), where optical excitation jointly modulates electronic band bending, FE switching, and Cu$^+$ ionic relaxation. The use of correlated Kelvin probe force microscopy, piezoresponse force microscopy, and conductive atomic force microscopy under above-bandgap illumination, reveals illumination-induced enhancement of surface work function, persistent surface photovoltage, reduced coercive field, and positive imprint shifts. These effects arise from synergistic photocarrier redistribution and slow Cu$^+$ migration that reshape interfacial depletion widths and internal electric fields. Illumination-assisted barrier lowering further enhances carrier injection and produces sweep-rate–dependent ferroionic transport hysteresis. Our results establish photoferroionic coupling as the governing mechanism for light-controlled band modulation and polarization stability in CIPS, providing a nanoscale framework for designing light-addressable FE memories, optoelectronic switches, and neuromorphic devices based on layered ferroionic materials.

# 1. Introduction

Two-dimensional (2D) ferroelectric (FE) semiconductors provide a unique materials platform in which spontaneous polarization, electronic transport, and optical excitation are intrinsically coupled at the nanoscale.[1] In these systems, photogenerated carriers interact strongly with polarization-bound charges and interfacial electric fields, enabling optical control over band bending, domain stability, and switching behavior.[2–7] Compared with conventional bulk FEs, reduced dielectric screening and atomically sharp interfaces in van der Waals (vdW) materials significantly amplify these effects, rendering their polarization states exceptionally sensitive to optical perturbation.[8–12] As a result, vdW FEs have attracted considerable interest for light-programmable memories,[7,13–15] optoelectronic switches,[16–23] tunable photodetectors,[24] and neuromorphic hardwares.[25]

Among vdW FEs, $CuInP_2S_6$ (CIPS) is distinguished by its robust room-temperature FE polarization, moderate bandgap (~2.8 eV), and intrinsic ferroionic character arising from the displacement and migration of mobile $Cu^+$ cations within the layered lattice.[26–28] Owing to this mixed electronic–ionic polarization mechanism, optical excitation in CIPS does not merely generate electronic photocarriers but can simultaneously activate slow ionic relaxation, giving rise to a coupled photoferroionic response. Macroscopic studies have reported illumination-induced changes in polarization,[29,30] surface photovoltage,[31] and electrical transport[18,32,33] in CIPS; however, the nanoscale mechanisms linking photocarrier redistribution, $Cu^+$ migration, and polarization-controlled interfacial band bending remain unresolved. In particular, direct experimental evidence correlating light-induced surface potential modulation with FE switching and ionic dynamics at metal/CIPS interfaces is still lacking.

In metal/FE heterostructures, polarization-bound charges, Schottky barriers, and interfacial trap states collectively govern the local electrostatic landscape and charge injection behavior. In CIPS, this framework is further complicated by the mobility of $Cu^+$ ions, which can drift under internal polarization fields as well as under external electrical or optical excitation.[28,29] Consequently, Cu$^+$ migration contributes not only to FE switching but also to the dynamic modification of depletion widths, band offsets, and interfacial charge compensation. Polarization evolution in CIPS therefore reflects a coupled electron–ion–dipole response rather than purely electronic screening, particularly under nonequilibrium conditions induced by illumination.

Despite extensive macroscopic investigations of photoinduced polarization and transport in CIPS,[18,29-34] nanoscale insight into how optical excitation reshapes interfacial band bending and switching energetics remains limited. Spatial averaging in bulk or planar device measurements obscures local variations in surface potential, polarization screening, and ionic redistribution that critically influence coercive field ($E_c$), imprint, retention, and carrier injection. Resolving these coupled processes requires nanoscale probes capable of simultaneously correlating surface electrostatics, FE order, and charge transport under optical excitation. Moreover, in nanoscale devices the top electrode can efficiently extract photogenerated carriers, enhancing external quantum efficiency and revealing spatially heterogeneous surface photoresponses that remain invisible in bulk measurements.[35]

Here, we employ correlated Kelvin Probe Force Microscopy (KPFM), Piezoresponse Force Microscopy (PFM), and Conductive Atomic Force Microscopy (C-AFM) to directly visualize photoferroionic coupling in exfoliated CIPS vdW heterostructures. The tracking of illumination-induced changes in surface work function ($\phi_w$), polarization switching behavior, and local current transport within the same device geometry reveals persistent surface photovoltage, reduced $E_c$, and positive imprint shifts arising from synergistic photocarrier redistribution and Cu$^+$ ferroionic relaxation. These findings provide direct nanoscale evidence of photoferroionic coupling at metal/CIPS interfaces and establish a mechanistic framework for light-controlled band modulation and polarization stability in vdW ferroionic FEs.

## 2. Experimental section

**Material synthesis and crystal growth:** Single crystals of CIPS were grown by chemical vapor transport using iodine ($I_2$) as the transport agent. A stoichiometric mixture of Cu (99.9%, Thermo Scientific), In (99.97%, Thermo Scientific), P (99%, Beantown Chemical) and S (99.99%, Thermo Scientific) powders, with a total mass of approximately 1.4 g, was combined with 40 mg of $I_2$ (99.8%, Thermo Scientific) and sealed under vacuum (~$10^{-5}$ Torr) in a quartz ampoule using an oxygen–acetylene torch. The transport reaction was carried out for one week in a two-zone furnace with a temperature gradient from 750°C (source) to 650°C (sink). After the growth, millimeter-sized, green plate-like single crystals with flat surfaces were obtained.

**Characterization of the CIPS single crystal:** The phase purity and crystallinity of the as-grown CIPS single crystals were verified by $\theta$-$2\theta$ X-ray diffraction measurements in a Rigaku Miniflex diffractometer equipped with a copper $k_\alpha$ source together with Raman spectroscopy measurements using the Horiba XploRA PLUS spectrometer.

**Exfoliation and device fabrication:** Thin CIPS flakes were mechanically exfoliated from bulk crystals using adhesive tape (Nitto SPV224 PVC) and transferred onto a PtSi-coated Si substrate. The PtSi layer was created by annealing a $\approx$ 20 nm thick Pt film deposited by sputtering on a heavily doped Si wafer devoid of any thermal oxide *insitu* at 200 ºC for 2 hrs. The exfoliated CIPS flakes were identified using optical microscopy and AFM to confirm their thickness and subsequently transferred onto the PtSi film, which served as the bottom electrode for electrical measurements. The asymmetric contact geometry between the PtSi and the metal (Au and Pt) coated AFM tip was carefully maintained to induce interfacial potential asymmetry, which is essential for probing band bending and polarization coupling phenomena.

**Scanning Probe Microscopy (SPM) measurements:** A Park systems (NX10) AFM system was used for the AFM, KPFM, PFM, and C-AFM measurements under ambient conditions (~25 °C, 40–50% relative humidity). The KPFM was performed in amplitude-modulated mode using Au-coated conductive tips to map the surface potential under the dark and illuminated conditions. The C-AFM based current-voltage characteristics were measured using the Pt-coated conductive probes (radius 10 nm) with a typical force constant of 18 N/m by scanning the voltage between ±10 volts. The FE and piezoelectric switching behaviors were studied in the vertical PFM (VPFM) and lateral PFM (LPFM) mode using Pt coated conductive tip (radius 10 nm). The PFM phase and amplitude signals were recorded under an AC excitation voltage of 1 V superimposed on a DC bias sweep of ±10 V. For thickness-dependent measurements, multiple thickness of flakes were analysed under identical conditions to extract $E_c$ and imprint from phase hysteresis loops. All measurements were repeated several times to confirm reproducibility, and the datasets were processed using XEI data processing and analysis software.

**Optical setup:** A continuous wave blue laser diode ($\lambda$ = 445 nm) with a 2 mm diameter spot size was used as the excitation source to illuminate the samples. The optical power density on the sample surface was varied from 0 to 150 mW cm$^{-2}$ by varying the diode current in conjunction with the use of neutral density (ND) filters. The light intensity on the sample was low enough to ensure minimal photothermal artifacts in the signal. Experiments were conducted on the bare PtSi surface under identical illumination conditions, yielding no measurable changes in surface potential, hysteresis behavior, or current response, ruling out laser-induced modification of the tip or tip–substrate interaction.

**Photoresponse measurement in FET device:** Single crystals of CIPS were mechanically exfoliated for device fabrication on a Si/SiO$_2$ (200 nm) substrate. Symmetric gold electrodes of 50 nm thickness were deposited via thermal evaporation using shadow masking, forming a metal–FE–metal configuration with in-plane carrier transport through a channel width of 25 μm, fabricated using the same diameter of tungsten wire. The back surface of the Si wafer is used as the back gate connection. The electrical measurements of the 2D CIPS device were performed using a two-channel Keithley (2602B) source-measure unit. The photoresponse of the device was recorded under the illumination of blue, green, and infrared diode lasers (445, 532, and 850 nm) across a broad temperature range in a cryogenic probe station.

## 3. Results and discussion

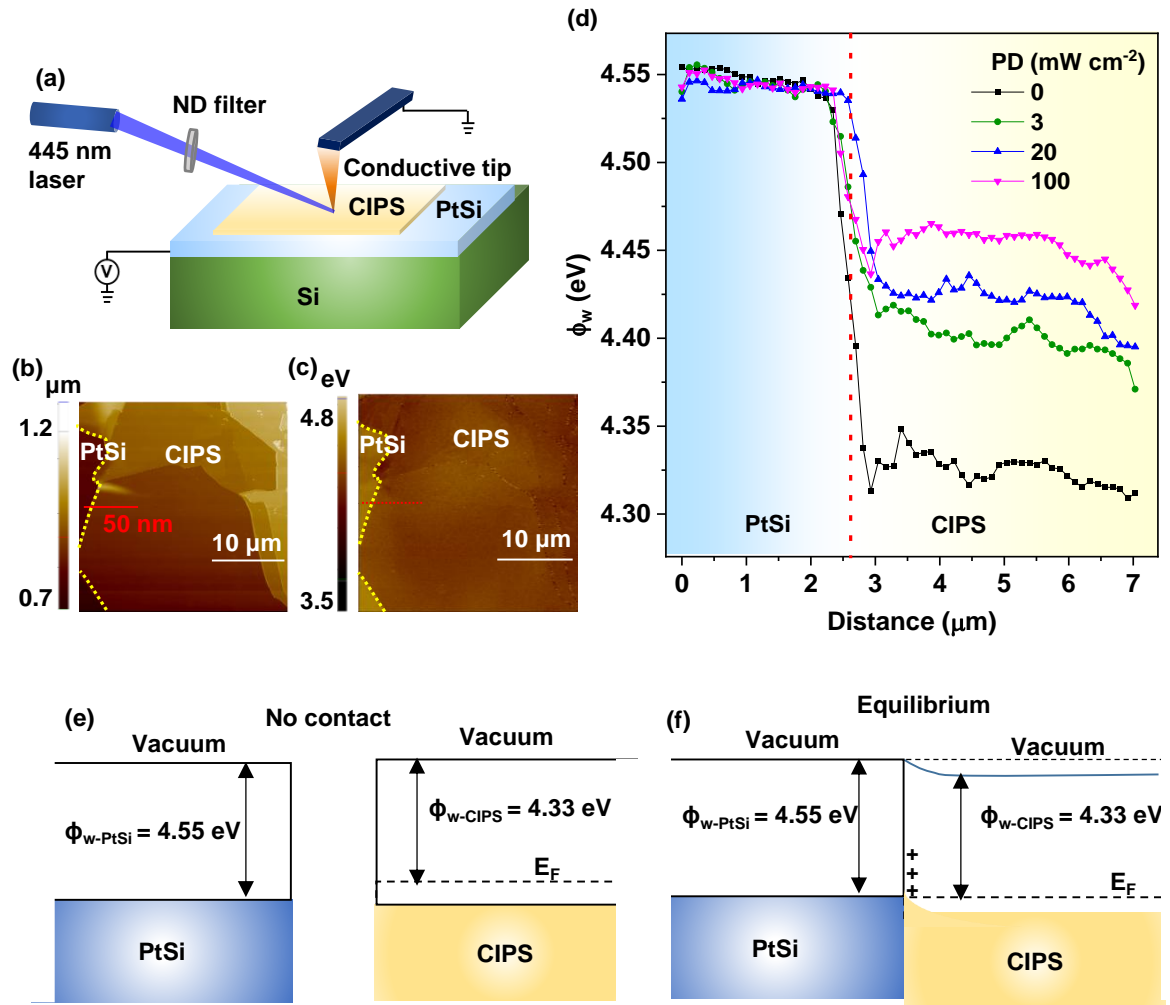

**Fig. 1:** (a) Schematic representation of the CIPS/PtSi/Si heterostructure used for NLPS photoresponse measurements. A conductive AFM tip was employed for KPFM, C-AFM, and PFM measurements. A blue diode laser ($\lambda$ = 445 nm), along with an ND filter, excites the sample with a variable flux of photons, (b) AFM topography and (c) corresponding KPFM $\phi_w$ maps of an exfoliated CIPS flake partially overlapping the PtSi bottom electrode. The yellow dotted lines indicate the boundary between CIPS and the adjacent PtSi layer. (d) KPFM $\phi_w$ line

profiles for both PtSi and CIPS layer measured across the red dotted line shown in (c), under dark and illuminated conditions. CIPS surface $\phi_w$ is increasing with increasing PD. (e,f) Schematic diagrams of the energy band alignment and charge transfer processes at the PtSi/CIPS heterojunction; (e) Energy levels of pristine PtSi and CIPS before contact, and (f) upward band bending at the interface after making the contact due to Fermi level equilibration between PtSi and CIPS.

Nanoscale local probe studies (NLPS) were performed on exfoliated CIPS crystals following confirmation of their phase purity and crystallinity via Raman spectroscopy and X-ray diffraction, as detailed in **Fig. S1** of the Supporting Information. For NLPS measurements, the exfoliated CIPS flakes were transferred onto PtSi-coated Si substrates to form well-defined CIPS/PtSi heterostructures. The corresponding measurement outcomes, obtained under different operational modes of the AFM system, are presented in the subsequent sections.

**3.1. Light modulated surface charge and polarization**

**Fig. 1a** represents the schematic of the experimental setup for integrated KPFM, PFM, and C-AFM measurements under controlled illumination using a 445 nm wavelength diode laser (details of the optical setup are explained in the Experimental section). This configuration enables simultaneous nanoscale mapping of surface potential, polarization state, and local current, along with surface topography, as a function of illumination intensity. For electrical measurements on flakes of varying thickness, the bias voltage was applied to the underlying PtSi electrode, while the conductive AFM tip serving as the top electrode was maintained at ground potential. **Figs. 1b,c** present the surface topography and the corresponding KPFM $\phi_w$ maps of a multilayer CIPS flake of spatially varying thickness positioned on the PtSi bottom electrode. The CIPS/PtSi boundary is marked by the yellow dotted line, as shown in the figures.

As evident from the topography in **Fig. 1b**, the color contrast defines changes in flake thickness. A clear variation in the local $\phi_w$ between the CIPS and PtSi regions is evident in **Fig. 1c**. The color scale indicates that the PtSi side exhibits a higher $\phi_w$ compared to the CIPS layers. For quantitative analysis, a ~50 nm-thick layer spanning the CIPS/PtSi interface was selected for KPFM line-profile measurements under dark conditions. The extracted $\phi_w$ values were referenced to the work function of the Au coated AFM tip ($\phi_{w\text{-tip}} \simeq 5.1$ eV), calibrated using a highly ordered pyrolytic graphite (HOPG) standard sample (see **Fig. S2**, Supporting Information). **Fig. 1d** shows the KPFM $\phi_w$ line profile acquired along the red dotted line across the CIPS/PtSi boundary shown in **Fig. 1c** under illumination power densities (PD) ranging from 0 to 100 mW cm$^{-2}$. In the dark (PD = 0 mW cm$^{-2}$), a distinct step in potential is observed, with the $\phi_{w\text{-PtSi}} \simeq 4.55$ eV exceeding that of CIPS ($\phi_{w\text{-CIPS}} \simeq 4.33$ eV). Upon illumination, the $\phi_{w\text{-CIPS}}$ increases monotonically as the PD is raised from 0 to 100 mW cm$^{-2}$, reflecting a pronounced photoresponse and interfacial charge redistribution. The KPFM $\phi_w$ maps measured at different PD (0–100 mW cm$^{-2}$) are shown in **Fig. S3** (Supporting Information), confirming that illumination progressively enhances the $\phi_{w\text{-CIPS}}$ while leaving the PtSi region largely unchanged. This trend is clearly captured in the line-profile data in **Fig. 1d**. The $\phi_w$ contrast obtained from the KPFM measurements offers direct insight into the electronic band alignment at the CIPS/PtSi junction, as illustrated in **Fig. 1e**. Under dark conditions, the $\phi_w$ offset ($\Delta\phi_w$) of $\simeq 0.22$ eV between the CIPS and PtSi layers produces upward band bending within CIPS near the interface. At equilibrium, this band bending creates an electron depleted region in CIPS side of the interface, giving rise to a built-in electric field directed from the CIPS layer toward the PtSi underlayer (**Fig. 1f**).

To investigate the influence of optical excitation on the electrostatic landscape of CIPS, we conducted KPFM measurements on a $\approx$ 40 nm-thick flake exhibiting an initially uniform contact potential ($V_{CPD}$) distribution. The $V_{CPD}$, defined as

$$V_{CPD} = \phi_{w\text{-tip}} - \phi_{w\text{-sample}}/e \qquad (1)$$

was recorded under progressively increasing PD from 0 to 150 mW cm$^{-2}$. Here $e$ is the electron charge.

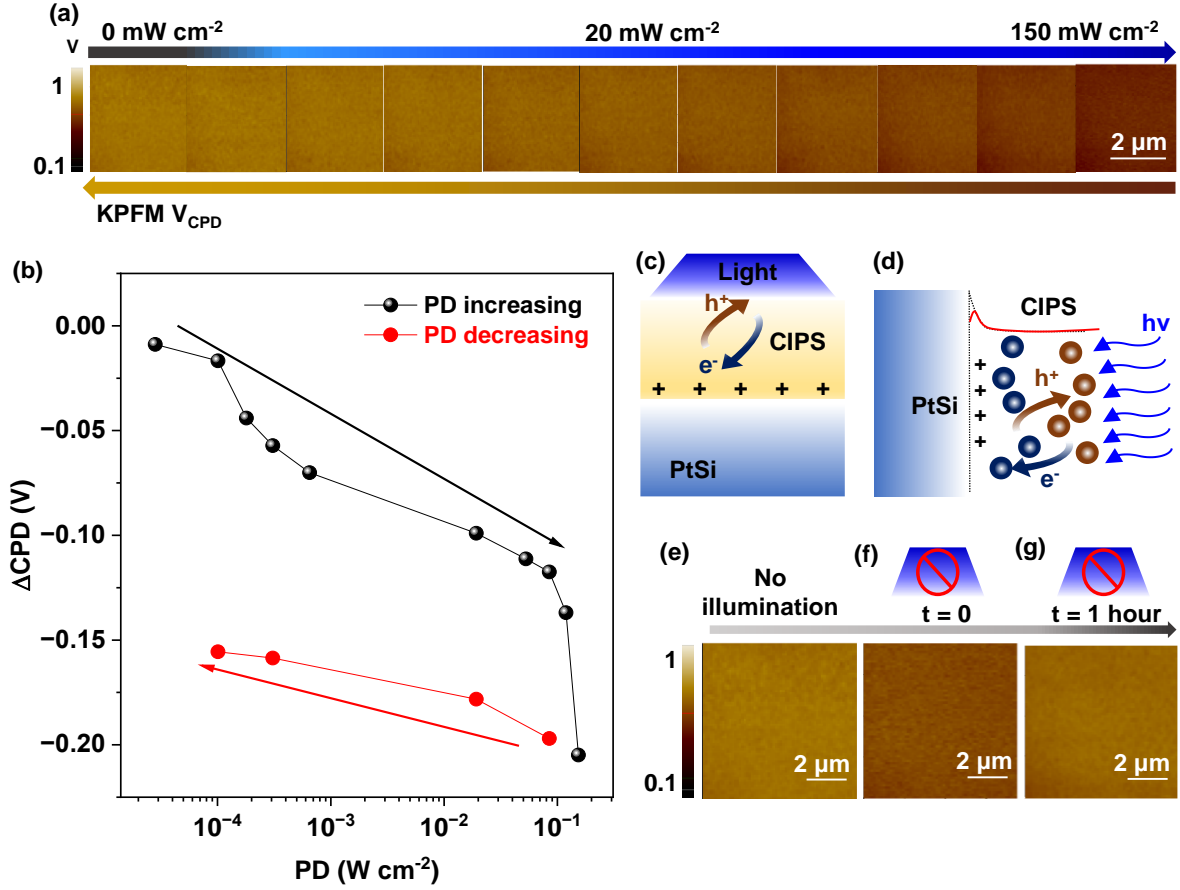

**Fig. 2:** (a) KPFM surface $V_{CPD}$ maps of a CIPS flake under varying PD from 0 to 150 mW cm$^{-2}$, showing a systematic decrease in surface potential with increasing PD. (b) Plot of the ΔCPD change as a function of PD during increasing (black) and decreasing (red) illumination. (c) Schematic illustration of the CIPS/PtSi interface, where an internal electric field forms a depletion region, giving rise to a Schottky-type junction. Upon illumination, photoexcited electron–hole pairs are generated within CIPS, with the built-in field driving electrons toward the PtSi/CIPS interface and holes toward the CIPS surface. (d) Photogenerated charge redistribution within CIPS: electron transport toward the depletion region reduces the potential barrier at the CIPS/PtSi interface, while holes accumulate at the CIPS surface, resulting in an enhanced relative $\phi_w$. (e-g) Temporal evolution of KPFM $V_{CPD}$ of CIPS: (e) in the dark, (f) immediately after switching off illumination ($t = 0$), and (g) after one hour in the dark ($t = 1$ hour).

Since the Au-coated KPFM tip exhibits a stable $\phi_w$ under light exposure, variations in $V_{CPD}$ unambiguously reflect changes in the sample work function, $\phi_{w-CIPS}$.[36,37] **Fig. 2a** shows a systematic decrease in $V_{CPD}$ with increasing PD, manifested as a progressive change in image contrast. Quantitatively, the illumination-induced shift in surface potential, ΔCPD, is defined as

$$\Delta CPD = (V_{CPD-light} - V_{CPD-dark}) \tag{2}$$

becomes increasingly negative with the higher PD (**Fig. 2b**), implying an upward shift in $\phi_{w-CIPS}$. Here, $V_{CPD-light}$ and $V_{CPD-dark}$ are the surface potentials measured in the light illumination and under dark conditions, respectively. This behaviour is consistent with photocarrier generation under 2.79 eV laser excitation ($\lambda = 445$ nm), followed by charge separation driven by internal electrostatic fields arising from interfacial band bending and intrinsic FE polarization in CIPS[4]. Such light-induced $\phi_w$ modulations are characteristic of semiconductors experiencing surface photovoltage (SPV) effects under above-bandgap excitation.[38] As KPFM is measured in non-contact mode, the tip- sample interaction is purely electrostatic, not a real electrical contact, and the tip does not inject any carrier during measurements. Hence, only the CIPS/PtSi interfacial band bending plays the driving role for carrier migration through CIPS. The CIPS side of the interface is electron depleted, as shown in **Fig. 2c**. Upon illumination, photogenerated electrons drift toward PtSi interface, while holes accumulate near the CIPS surface, producing a SPV. As illumination increases, the accumulation of photogenerated holes at the upper surface enhances the measured $\phi_w$, thereby lowering $V_{CPD}$. The CIPS/PtSi interfacial potential barrier also decreases with increasing PD (**Fig. 2d**). This monotonic trend in $\Delta_{CPD}$ (**Fig. 2b**) thus reflects increasingly efficient photocarrier separation and surface charging at higher optical flux, in line with reported SPV behavior in 2D FEs.[39,40]

The dependence of $\Delta_{CPD}$ on PD is quasi-linear up to ~100 mW cm$^{-2}$, indicative of a steady-state photocarrier population governed by generation–recombination balance. Beyond this threshold, however, $\phi_{w\text{-CIPS}}$ exhibits a sharp, irreversible increase, suggesting a local non-equilibrium transition within the material, possibly associated with defect charging, ionic displacement, or light-activated polarization modification—phenomena observed in related vdW FEs under strong illumination.[41-43] The irreversibility upon reducing PD (red arrow in **Fig. 2b**) further supports the presence of metastable charge configurations. To probe this behavior, we examined the temporal relaxation of the surface potential under moderate illumination. **Figs. 2e–g** show sequential KPFM images recorded in the dark (**Fig. 2e**), after continuous illumination with moderate PD (<10 mW cm$^{-2}$) for 30 minutes, followed by immediate light-off ($t = 0$) and after one hour of the off-illumination condition ($t = 60$ minutes), as shown in **Figs. 2f** and **2g**, respectively. From the color contrast variation, it is clearly noticeable that the surface potential immediately after light removal (**Figs. 2f**) is not the same as the one measured in pristine dark condition (**Figs. 2e**), indicating the presence of persistent surface photovoltage (PSPV). The gradual but incomplete recovery over one hour (**Figs. 2g**) indicates slow migration of trapped photocarriers. Such PSPV effects are commonly attributed to defect-mediated trapping, space-charge accumulation, or slow ionic migration—particularly involving chalcogen vacancies or cation non-stoichiometry in layered sulphides.[44-46] In the present case, trapped holes at the surface and trapped electrons near the CIPS/PtSi interface sustain the potential gradient long after illumination ceases, consistent with previous reports of persistent photoconductivity and photovoltage in defect-rich 2D semiconductors.[47-49]

To establish a direct correlation between the photoinduced potential variation and the FE response of CIPS, the temporal evolution of the PFM phase was examined under dark and illuminated conditions. During these measurements, a low and constant tip-loading force was maintained to minimize mechanically driven domain switching through the flexoelectric effect, as demonstrated by Zhang et al.[50]

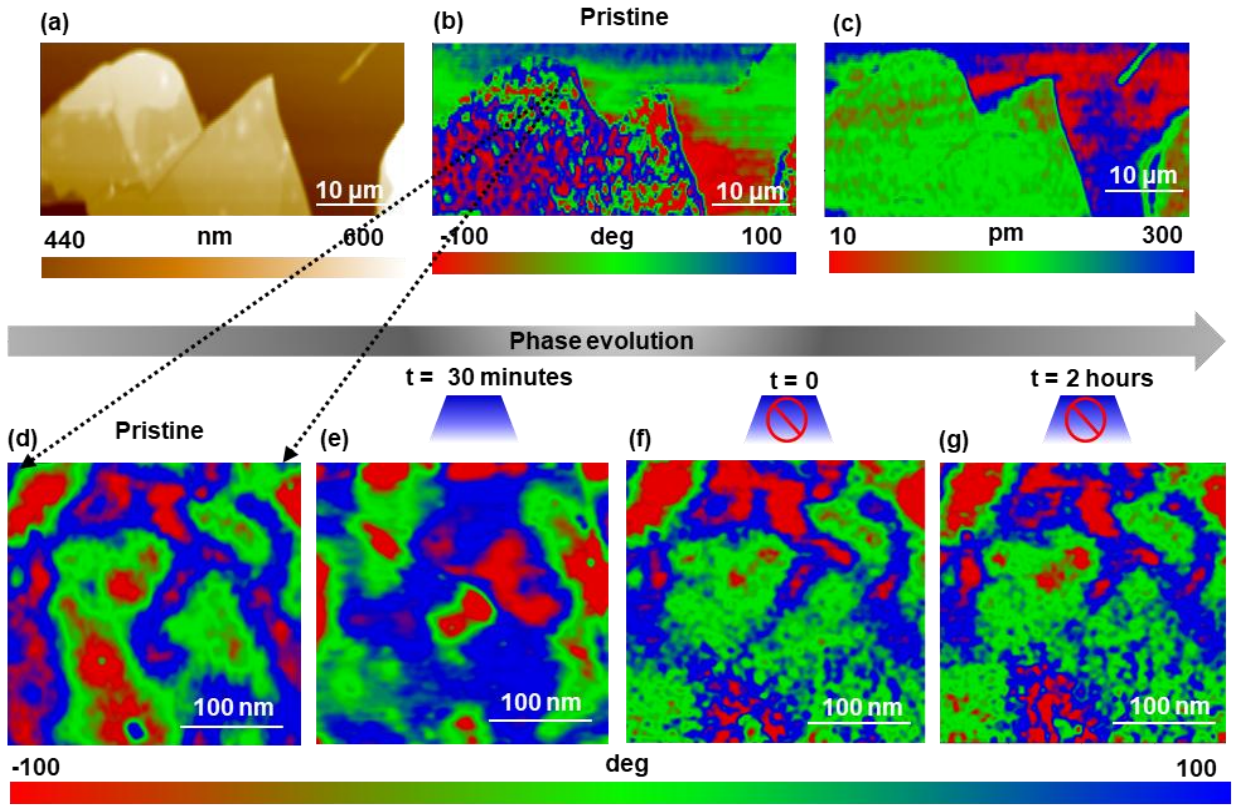

**Fig. 3:** (a) AFM topography, (b) PFM phase and (c) PFM amplitude of pristine layered CIPS flakes. Temporal evolution of the PFM phase images of selected region of pristine CIPS flake before and after illumination. (d) Pristine CIPS, (e) 30 minutes after continuous light illumination, (f) immediately after the light is switched off, and (g) 2 hours after switching off the light.

The AC drive voltage and lock-in settings for were also kept constant to avoid artifacts in electromechanical response during all the measurements. **Figs. 3a-c** show the AFM topography together with the PFM phase and amplitude image of a pristine layered CIPS nanoflake. Randomly oriented FE domains are visible in both the phase and amplitude images, as indicated by the color contrast bar in **Figs. 3b** and **3c**.[51] A region with ~50 nm thickness (around the dotted black arrow in **Fig. 3b**) was selected for detailed time-dependent phase measurements under different illumination conditions. In the dark (**Fig. 3d**), the pristine CIPS flake exhibits PFM phase contrast with multiple, randomly oriented FE domains. Upon continuous illumination for 30 minutes at a PD ≈ 100 mW cm$^{-2}$ (**Fig. 3e**), a pronounced modification in the phase contrast is observed. This modification reflects redistribution of internal electric fields induced by photogenerated carriers, indicating that illumination perturbs the local polarization configuration. Notably, light-induced domain switching in CIPS has previously been reported under 405 nm excitation at higher laser intensities, attributed to Cu$^+$ ion migration.[29] Our present results reveal partial FE domain alteration at a longer wavelength (445 nm) and substantially lower optical PD, highlighting the sensitivity of CIPS to carrier-mediated optical modulation. After the illumination is switched off, the modified domain pattern does not immediately revert to its pristine configuration (**Fig. 3f**). Instead, the altered PFM phase distribution persists for an extended duration and gradually relaxes toward the initial state only over prolonged timescales (**Figs. 3g**). The corresponding amplitude images are provided in **Fig. S4** (Supporting Information). This behaviour aligns with the persistent shifts observed in the KPFM measurements (**Fig. 2**) and suggests that photoinduced carriers remain trapped, facilitating partial polarization screening even in the absence of continuing excitation. The observed behavior is further implicated by interfacial band bending at the conductive AFM tip/CIPS/PtSi stack as a critical factor governing the measured surface FE response. To evaluate this interplay, KPFM, PFM, and C-AFM measurements were performed both in the dark and under illumination while applying a controlled DC bias. These measurements confirm that the observed photoresponse arises from the combined influence of polarization-dependent band bending and light-induced carrier dynamics, which together define the electrostatic behaviour of the CIPS/metal heterostructure.

### 3.2. KPFM measurements under spatial poling

**Figs. 4a,b** present the $V_{CPD}$ maps obtained from a ≈ 50 nm-thick CIPS flake immediately after spatial poling ($t = 0$) and after 24 hours of relaxation following removal of the external electric field ($E_{ext}$), respectively. The corresponding line-profile averages of the whole region (**Fig. 4c**) show a pronounced potential contrast between oppositely written domains (±10 V and ±5 V), confirming reliable FE polarization reversal within the electrically patterned regions. Although the contrast decreases after the relaxation, a measurable $V_{CPD}$ difference remains (**Fig. 4b**), evidencing long-lived retention and incomplete compensation of polarization charges. The gradual decay of the potential contrast is attributed to progressive screening of the depolarization field by mobile charge carriers, defect states, and slow ionic motion in CIPS. In layered CIPS, $Cu^+$ cations exhibit weak coordination and can migrate under internal fields, redistributing space charges near the interface and reducing the polarization-induced band bending over time. This behavior is consistent with previously reported ionic relaxation in copper-based van der Waals FEs.[28,52-54] The spatial variation in $V_{CPD}$ thus reflects the local modulation of the surface $\phi_w$ arising from polarization-dependent surface charge modulation in CIPS. To understand the correlation between the $V_{CPD}$ and surface polarization at different poled regions, one must understand the two steps involved in this procedure: (i) Domain writing by FE switching under an applied DC electric field ($E_{ext}$) with the tip in contact mode; and (ii) $V_{CPD}$ imaging in non-contact KPFM after removing $E_{ext}$, as described in the following section.

i) Domain Writing ($E_{ext} \neq 0$, Tip in Contact Mode): The domain-writing process is shown in **Figs. 4d–f**. Different DC voltages were applied to the PtSi bottom electrode while the Au tip was grounded. In the absence of an applied bias, dipoles in the CIPS flake remain randomly oriented, producing negligible net polarization (**Fig. 4d**). A positive $E_{ext}$ aligns the polarization ($P_{up}$) from the PtSi substrate toward the Au tip (**Fig. 4e**), with the polarization magnitude increasing with the applied voltage. Here, the red and green arrows indicate the directions of the $E_{ext}$ and polarization, respectively. Under negative $E_{ext}$, the polarization reverses direction ($P_{down}$), pointing from the top Au side toward the PtSi electrode (**Fig. 4f**). Once $E_{ext}$ is removed, these written polarization states are retained, and the potential barrier at CIPS/PtSi interface is changed due to the intrinsic FE retention. (**Figs. 4g–i**). The barrier height increases when polarization is away from the interface ($P_{up}$), while it decreases for polarization towards the interface ($P_{down}$). Ferroelectric polarization introduces an internal electric field that partially compensates the built-in Schottky field at the interface along the polarization direction, thereby weakening band bending and reducing the effective barrier height.[55] ii) $V_{CPD}$ Imaging after Field Removal ($E_{ext} = 0$): Following domain writing, the $V_{CPD}$ was measured in non-contact KPFM using a small AC voltage drive (1 V and 17 kHz frequency). In absence of $E_{ext}$ during imaging, the spatial contrast in **Fig. 4a** originates solely from the remanent FE polarization, and the polarization mediated CIPS/PtSi Schottky barriers. In unpoled regions, $V_{CPD}$ appears spatially irregular, reflecting the coexistence of preferred or mixed polarization orientations. Positively poled regions generate positive unscreened bound charge at the CIPS surface, increasing the local $\phi_w$ and producing a lower $V_{CPD}$, in agreement with the contrast observed in **Fig. 4c**. Conversely, negatively poled domains accumulate negative bound charge, which decreases $\phi_w$ and results in a higher $V_{CPD}$ (**Fig. 4c**). The magnitude of this shift increases systematically with the applied poling voltage.

The KPFM measurements were repeated on a second ≈ 50 nm-thick CIPS flake using the same spatial poling procedure, followed by monitoring the temporal evolution of $V_{CPD}$ first in the dark and subsequently under optical illumination (**Fig. S5** in Supporting Information).

In the dark, each poled region exhibits a clear $V_{CPD}$ contrast comparable to that in **Fig. 4a**, directly reflecting the polarization state of the pre-poled regions.

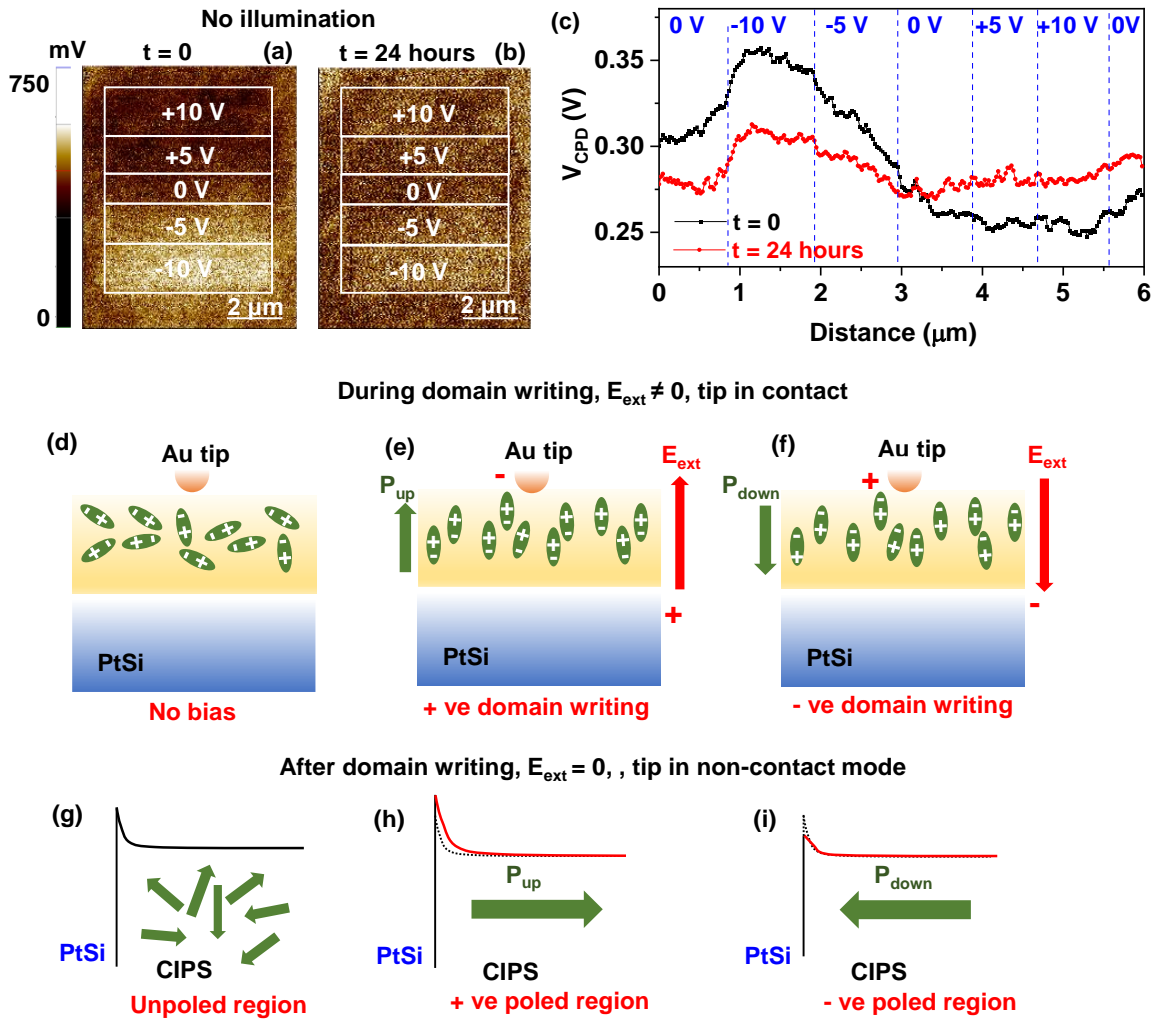

**Fig. 4**: KPFM $V_{CPD}$ of a CIPS flake after local poling by applying DC sample bias without any illumination exposure. (a) $V_{CPD}$ image acquired immediately after poling, and (b) 24 hours after poling. (c) Average line profile of $V_{CPD}$ calculated from the entire region of (a) (black) and (b) (red). (d-f): Domain writing process using Au-coated tip by applying (d) no bias, (e) + ve bias, and (f) - ve bias voltage. Applied field direction $E_{ext}$ is shown by the red arrow. The green arrow shows the polarization direction. The green-marked regions in CIPS indicate the dipolar orientation. (g-i): Modulation of CIPS/PtSi interfacial band bending in different pre-poled

regions removing the applying bias, (g) unpoled, (h) + ve poled, and (i) - ve poled regions, respectively.

Upon illumination, however, a remarkable reduction in the $V_{CPD}$ contrast between oppositely poled regions is observed. This reduction is attributed to photogenerated carriers that effectively screen the polarization-induced bound surface charges. When the illumination is switched off, the surface potential contrast partially recovers. This recovery indicates that photo-induced carriers recombine or diffuse away, allowing the depolarization field to re-establish. Nevertheless, the contrast does not fully return to its initial value, suggesting persistent screening contributions from trapped carriers or slow ionic migration within CIPS. The extent to which illumination alters the $V_{CPD}$ and the efficiency of local screening depends sensitively on the illumination intensity, trap density, and the underlying polarization state. Under stronger light exposure, enhanced photocarrier generation more effectively narrows the depletion region, producing a larger $V_{CPD}$ shift, while the intrinsic polarization further modulates interfacial barrier heights, adding asymmetry to the resulting surface potential landscape.

### 3.3. *I-V* measurements to understand dynamic switching and charge transport

To relate the static interfacial band bending with the dynamic charge transport, the current (*I*) – voltage (*V*) characteristics of the PtSi/CIPS (20 nm) /Pt junction have been measured. Here, Pt-coated conductive tip is used to measure local current in C-AFM mode. **Fig. 5a** shows the room-temperature I–V characteristics of the junction. The current was recorded during sequential voltage sweeps (path 1 → 6) from +10 V → −10 V (forward, red 1 → 3) and subsequently from −10 V → +10 V (reverse, blue 4 → 6). The measured current remains in the picoampere range throughout the ±10 V sweep, consistent with the high-resistance semiconducting nature of CIPS. The I–V response exhibits four key features: (i) rectification behavior ($I_{-10V}/I_{+10V} \approx 1.4$), (ii) pronounced hysteresis between the two sweeps, (iii) near symmetry of the forward and reverse curves, pointing to interface-limited transport, and (iv) a clear current cross-over at ≈ ±5 V, remarkably close to the coercive voltage ($V_c$) obtained from PFM phase loops of a CIPS flake of similar thickness (≈ 20 nm). These behaviors originate from the strong asymmetry between the two metal/CIPS interfaces. Using the Schottky–Mott approach,[56] the electron injection barriers derived from the work functions ($\phi_{w\text{-}PtSi} \approx 4.55$ eV, $\phi_{w\text{-}CIPS} \approx 4.33$ eV, this work) are ~0.22 eV at PtSi/CIPS and ~1.32 eV at Pt/CIPS ($\phi_{w\text{-}Pt} \approx 5.65$ eV[57]). This produces a highly asymmetric double-Schottky structure, with a nearly ohmic injection interface at PtSi and a strongly injection-limited interface at Pt. As a result, the electrostatic band bending is steep near the Pt tip and minimal near PtSi (**Fig. 5b**). For + ve bias on PtSi, electrons must be injected from the grounded Pt tip across the large 1.32 eV barrier, resulting in extremely small currents (**Fig. 5c**). For - ve bias, electrons originate from the PtSi contact across its small 0.22 eV barrier; although they still encounter the large Pt barrier, the strong tip-localized electric field and image-force lowering[58] increase the transmission probability (**Fig. 5d**), resulting in slightly larger currents in the – ve voltage region (**Fig. 5a**).

Beyond these static considerations, the dynamic ferroionic nature of CIPS plays a central role in shaping the hysteresis seen in the I–V characteristics. CIPS contains a small but mobile $Cu^+$ sublattice, and its spontaneous polarization is tied to the displacement of these ions.[28] Consequently, bias sweeps not only modulate the Schottky barriers but also drive polarization switching, slow ionic redistribution, and trap occupancy changes, all of which alter the internal electric field and effective barrier heights on the timescale of the voltage sweep. During the forward (+10 V → −10 V) sweep, the field gradually rotates the polarization from pointing toward the Pt tip at +10 V to pointing toward the PtSi bottom electrode at −10 V (**Fig. 5a**, path 1→3). - ve bias favors electron injection from PtSi and stabilizes a strong downward (Pt → PtSi) polarization state.

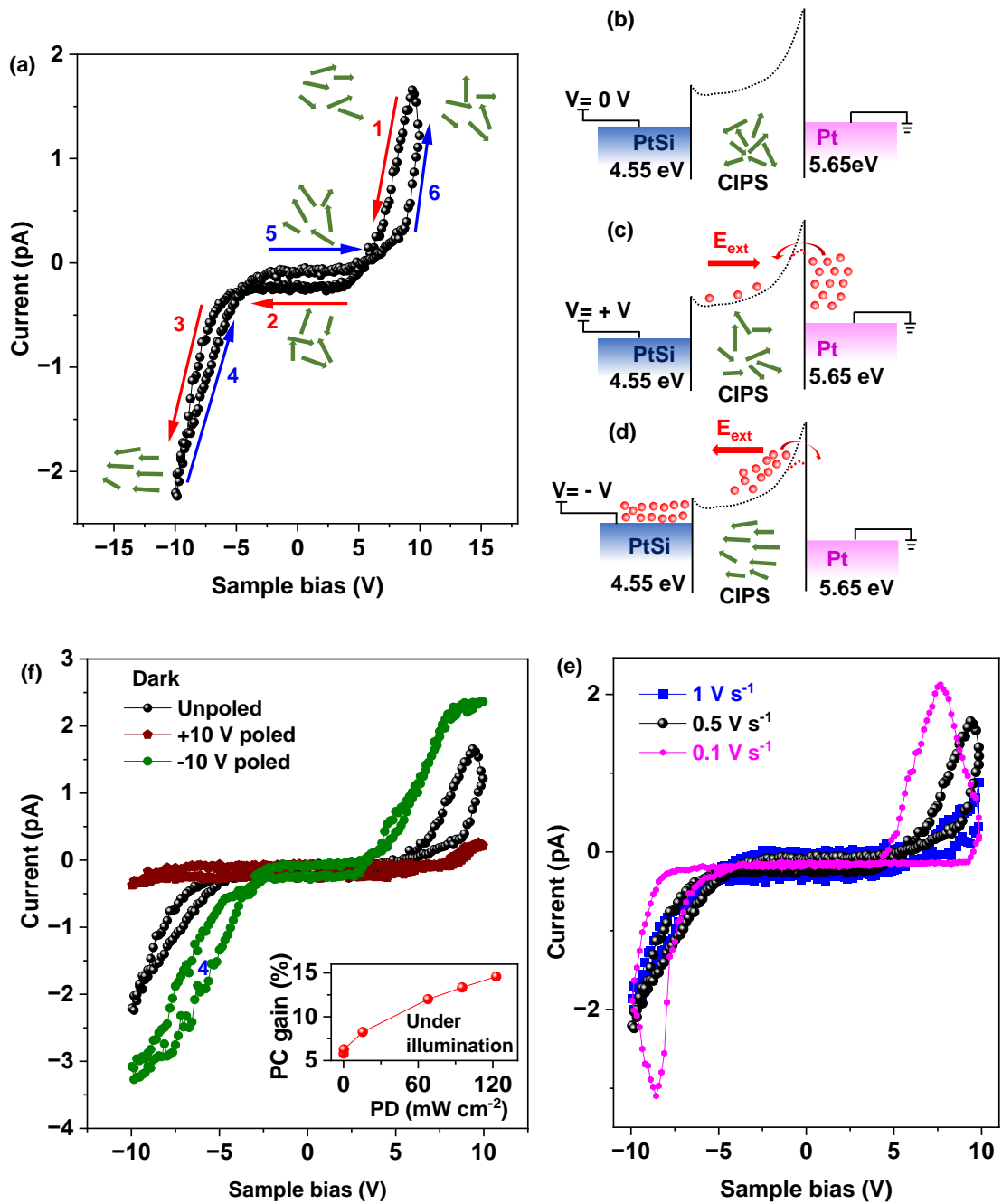

**Fig. 5:** C-AFM I–V spectroscopy of a 20 nm thick CIPS flake placed on a PtSi-coated Si substrate, revealing FE polarization–modulated conduction behavior. (a) The unpoled *I–V* curve measured between the +10 → −10 V (red arrows) and −10 → +10 V (blue arrows) sweep.

(b-e) Band bending at the interfaces under different voltage sweep: (b) when no voltage is applied, (c) under + ve and (d) under – ve voltage sweep. Red circles represents the electrons as charge carrier in CIPS. (e) *I-V* curves for an unpoled sample measured at different voltage sweep rates. (f) *I-V* curves for unpoled, + vely poled, and - vely poled samples at the same spot. The relative photocurrent at different illumination power densities is shown in the inset to (f).

This again lowers the effective Schottky barrier at PtSi[55] as well as the Pt tip (image-force lowering[58]), dramatically increasing the current once the polarization fully switches. Through this forward sweep, the $Cu^+$ ions within the layers start migrating, and this migration is opposite to the direction of the polarization. During the reverse sweep (−10 V → +10 V), the system begins in this downward-polarized state. As the voltage approaches 0 V, and then becomes + ve, the polarization resists switching due to remanent slow $Cu^+$ ionic displacement and deep electronic traps. Hence, for $|V| < V_c$, the internal dipolar field remains biased downward, continuing to lower the barrier and yielding a higher current than in the forward sweep (path 4→5 in **Fig. 5a**). Once $V > V_c$, the polarization abruptly flips upward, inverting the interfacial bound charge and sharply reconfiguring the Schottky barriers. The forward and reverse current paths intersect at the switching point, producing the observed cross-over. Additional processes, including, $Cu^+$ ion migration[54], field-assisted trapping/detrapping, and slow relaxation dynamics shift the precise position of the cross-over but do not dominate its origin. The close match between the cross-over voltage and the PFM $V_c$ confirms that FE switching is the primary mechanism modulating the transport nonlinearity. **Fig. 5e** shows the I–V characteristics of an unpoled CIPS sample measured at three different voltage sweep rates (0.1–1 V s$^{-1}$). Notably, the hysteresis window becomes wider at slower sweep rates, and the voltage cross-over shifts towards 0 V, which is indicative of a time-dependent ionic motion. Such behavior in CIPS, often observed as sweeping speed-controlled self-rectification governed by field-assisted Cu$^+$ migration, were reported in earlier studies.[28,54,59] A decreasing crossover in the *I-V* curve with slower voltage sweep rate (0.1 V s$^{-1}$) indicates that the system has sufficient time to reach a quasi-equilibrium configuration during the sweep. Overall, the sweep-rate dependence of the I–V curves demonstrates that slow Cu$^+$ ion migration plays a crucial role in shaping the dynamic electrical behavior of CIPS. The broader hysteresis at slow sweep rates provides direct evidence of strong coupling between ionic motion, space-charge accumulation, and the evolving internal electric field in this layered FE material.

**Fig. 5f** shows the I–V curves measured after applying different pre-poling voltages (0 V, –10 V, and +10 V). The −10 V pre-poled state exhibits the highest current. Negative poling aligns the polarization vector from the Pt electrode toward the PtSi side, which reduces the depletion width and lowers the effective Schottky barrier at the PtSi/CIPS interface. The resulting redistribution of the internal electric field further decreases the barrier at the Pt top contact via image-force–assisted barrier lowering,[60] thereby enhancing electron transmission and facilitating charge transfer across the device. Because this interface constitutes the dominant transport bottleneck, even a modest barrier reduction yields a substantial increase in current, which persists even after removal of the poling field due to polarization retention. Conversely, +10 V pre-poling generates positive bound charge at the Pt/CIPS interface, but the high effective potential barrier at that interface limits the charge transfer, as well as the polarization switching. The zero-volt pre-poled state, containing partially relaxed or mixed domains, yields an intermediate current level. The inset of **Fig. 5f** shows the variation of the Photocurrent (PC) gain, $((I_{photo}-I_{dark})/I_{dark}) \times 100\%$,[61] measured at +10 V bias voltage as a function of PD (the I-V curves measured at different PD are shown in **Fig. S6**, Supporting Information). The photocurrent increases monotonically with the rising illumination intensity, indicating a light-induced reduction of the interfacial barrier through screening and enhanced carrier generation. This photo-enhancement further corroborates the photoactive nature of the CIPS, where photogenerated carriers assist in screening the polarization field and modulating the local conduction pathways. The photoresponse of CIPS was further evaluated using a three-terminal field-effect transistor (FET) configuration (**Fig. S7**, Supporting Information), which provides additional confirmation of the bulk photovoltaic effect in CIPS. This light-tunable current response demonstrates the synergistic coupling of FE and photoelectric effects in CIPS, which is essential for nonvolatile photoferroelectric memory and optoelectronic switching applications.

**3.4. PFM measurement to investigate the photoferroelectric effect**

The PFM phase hysteresis loops are measured for CIPS flakes of different thickness under light illumination to directly probe the effect of photogenerated carriers on FE switching, interfacial band bending, and modulation of internal fields within the CIPS/PtSi heterostructure. Exfoliated CIPS flakes exhibit both in-plane and out-of-plane polarization components, down to ≈ 20 nm thickness, with a dominant out-of-plane polarization as shown by LPFM and VPFM measurements (**Fig. S8**, Supporting Information). Furthermore, VPFM measurements reveal a characteristic box-in-box FE domain switching pattern under reverse DC bias application. **Fig. 6a** and **6b** show the VPFM phase images of 50 nm and 20 nm thick CIPS flakes, respectively, after DC bias writing. A pronounced ~180⁰ phase contrast is observed between oppositely poled regions, confirming FE domain reversal.[62] The corresponding VPFM amplitude images are provided in **Fig. S9** (Supporting Information). **Fig. 6c** and **6d** show the evolution of the PFM phase hysteresis loops of CIPS flakes with different thicknesses (50 nm and 20 nm) as a function of laser intensity. Under dark conditions, both samples exhibit well-defined loops with clear switching at their respective $V_c$. The asymmetry between the two interfaces (PtSi/CIPS/Pt) produces a shift of the hysteresis loop along the voltage axis, with the imprint being larger for thinner flakes. The higher $V_c$ and larger imprint observed in thinner flake can be attributed to several thickness-dependent effects, such as enhanced depolarization fields,[63] reduced effective electric field beneath the PFM tip,[64] stronger substrate-induced mechanical clamping,[65] and increased interfacial pinning effects.[65] We observed that both the $V_c$ and the extracted $E_c$ (= $V_c$/thickness) increase as the flake thickness decreases from 50 nm to 20 nm. This thickness dependence is attributed to the asymmetric Pt/CIPS/PtSi contact geometry, where interface-controlled band bending and built-in internal fields exert a stronger influence on the overall potential distribution in thinner flakes, thereby increasing the effective switching field. As the PD of illumination is increased, the phase loops undergo systematic modification. **Figs. 6e** and **6f** shows the imprint voltage ($V_{imp}$) (left y-axis, black color) and $E_c$ (right y-axis, red color) as a function of PD for 50 nm and 20 nm flakes, calculated from **Figs. 6c,d**, respectively. At moderate PD (≤ 20 mW cm⁻²), the hysteresis loops progressively narrow, indicating a pronounced reduction in the $E_c$ for both 50 and 20 nm flakes. For PD > 20 mW cm⁻², $E_c$ reaches a plateau in the 50 nm thick flake, persisting up to PD ≈ 70 mW cm⁻². Higher illumination PDs were not explored for this flake, as prior KPFM measurements on flakes of same thickness showed irreversible changes in surface potential beyond this PD. In contrast, for the 20 nm thick flake, $E_c$ continues to decrease with increasing PD. Measurements were extended up to PD ≈ 150 mW cm⁻² for this flake, over which the decreasing trend remains unchanged. The $V_{imp}$ tends to shift towards the positive voltage side with the increasing PD.

These light-induced changes arise from the interplay between photocarrier generation, ionic polarization dynamics, and asymmetric Schottky barriers at the PtSi/CIPS/Pt junction, mechanisms that are fully consistent with the trends observed in the I–V characteristics (**Fig. 5**). In the absence of illumination, polarization switching in CIPS is strongly influenced by the asymmetric double-Schottky interface, with the Pt/CIPS contact presenting a large electron-injection barrier (**Fig. 5b**) and thus generating a strong tip-localized electric field during voltage sweeps. This asymmetry facilitates localized downward ($P_{down}$) polarization at negative bias while opposing upward polarization at positive bias, giving rise to the $V_{imp}$ observed in PFM measurements. The same directionality of voltage application (+10 V → −10 V; -10 V → +10 V) during the PFM and *I-V* measurements preserves this built-in asymmetry. However, upon illumination, photogenerated carriers are introduced at both interfaces and within the bulk of the CIPS. These carriers partially screen the bound polarization charges and reduce the internal depolarization field. Consequently, the effective field required to switch the $Cu^+$-driven FE dipoles decreases, leading to a reduced $E_c$ in the PFM loops measured under illumination. This photocarrier screening effect mirrors the behaviour observed in the *I–V* curves, where illumination lowers the effective Schottky barrier height and enhances carrier injection, manifested as light-enhanced current. At higher PDs, the response becomes thickness dependent. In the 50 nm flake, the density of photocarriers becomes sufficient to nearly saturate the available screening channels, leading to a plateau in $E_c$. Additionally, stronger bulk-like FE behavior and reduced sensitivity to interface-dominated effects limit further reduction in the switching field. In contrast, the response of the 20 nm flake remains dominated by surface and interface effects, with a larger fraction of polarization charges and trap states accessible to photoexcitation. As a result, continued photocarrier generation and redistribution further enhance screening and reduce domain pinning, allowing $E_c$ to decrease monotonically even at higher PDs. The illumination-induced shift of the $V_{imp}$ toward positive bias can be attributed to asymmetric photocarrier accumulation at the metal–FE interfaces, which modifies the internal built-in field and alters the balance of interfacial screening. Moreover, illumination alters the ferroionic landscape of CIPS. Photocarriers modulate the local electrostatic potential and facilitate redistribution of the mobile $Cu^+$ ions, which participate directly in the polarization switching. This redistribution tends to stabilize the $P_{down}$ state, similar to the enhanced conduction seen in the negative-bias region of the *I–V* measurements, resulting in the positive imprint. In essence, light-assisted Cu$^+$ migration and photocarrier-induced interfacial screening jointly shift the energy landscape, increasing the stability of the $P_{\text{down}}$ configuration.

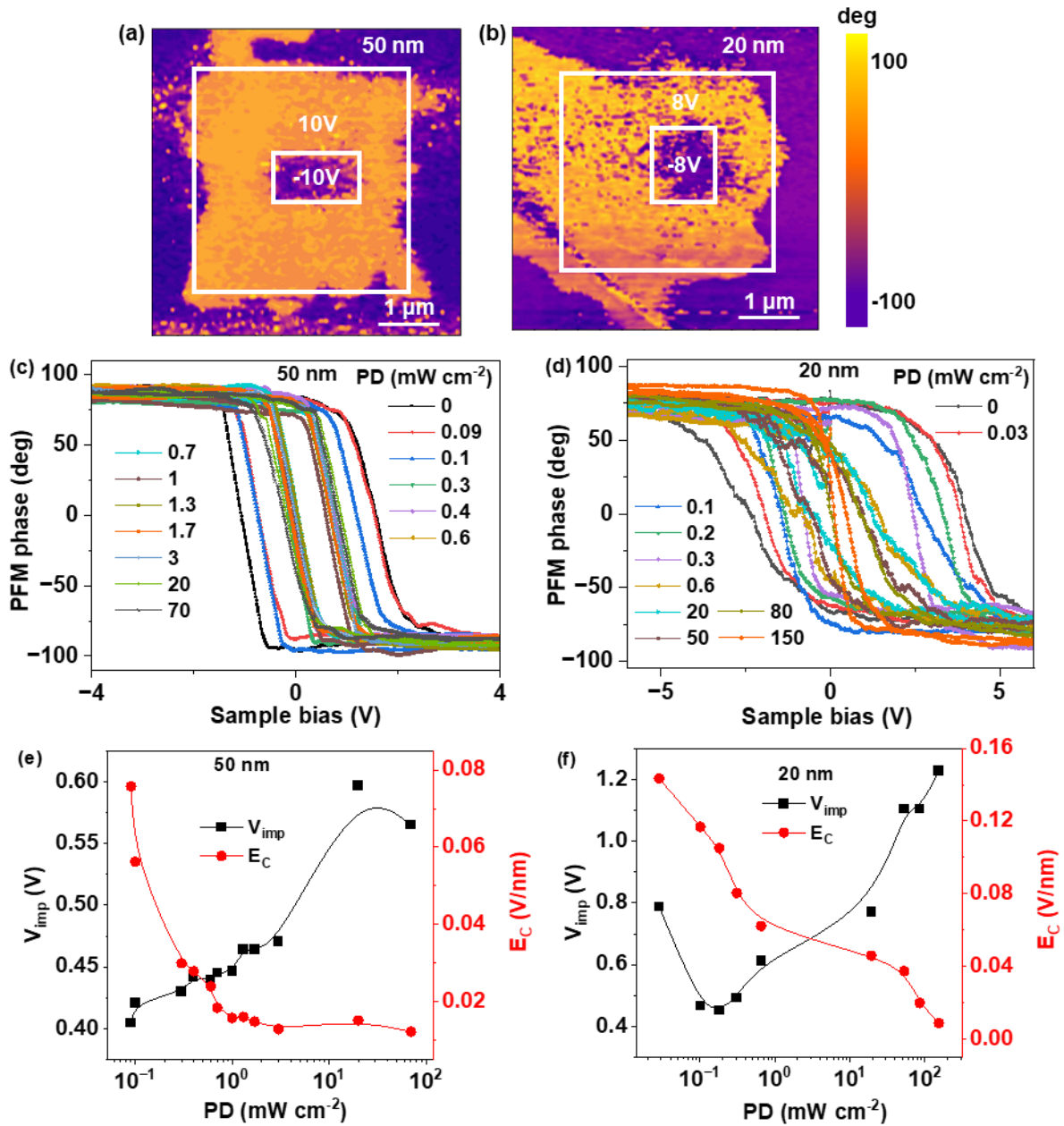

**Fig. 6:** FE polarization switching by VPFM for 50 nm and 20 nm thick CIPS flakes. PFM phase images for (a) 50 nm and (b) 20 nm CIPS with box-in-box written pattern by applying reverse DC bias. VPFM phase hysteresis measured in point spectroscopy mode at different PDs for (c) 50 nm and (d) 20 nm thick CIPS flakes. The measured $V_{\text{imp}}$ (black color, left y-axis) and $E_c$ (red color, right y-axis) for 50 nm and 20 nm thick samples are shown in (e) and (f), respectively.

Overall, the illumination-induced modulation of surface potential and transport can involve electronic screening and interface trapping; however, these effects alone do not fully account for the observations reported here. Specifically, the persistence of the surface photovoltage after light removal, the sweep-rate dependence of the transport hysteresis, and the stabilization of polarization under illumination. These behaviors are consistent with the involvement of slow ferroionic relaxation associated with mobile $Cu^+$ cations, acting in conjunction with photocarrier redistribution. The synergistic interplay of electronic and ionic responses provides a self-consistent explanation for the concurrent changes in interfacial band bending, coercive field, and polarization asymmetry observed under illumination.

## 4. Conclusion

In summary, we have demonstrated that the interplay between photocarriers, FE polarization, and ferroionic $Cu^+$ migration fundamentally governs light-tunable band modulation and switching behavior of the vdW FE CIPS. Through spatially resolved KPFM, we showed that illumination produces a substantial increase in the CIPS $\phi_w$, accompanied by a persistent surface photovoltage arising from photocarrier separation and long-lived trap occupancy. The resulting modification of interfacial band bending directly influences polarization state, as confirmed by illumination-dependent PFM hysteresis loops that exhibit reduced $E_c$ and a progressive positive imprint. These changes originate from the synergistic screening of polarization-bound charges by photocarriers and light-assisted redistribution of mobile $Cu^+$ ions, which together reshape depletion widths and internal electrostatic fields. C-AFM $I$–$V$ spectroscopy reveals the consequences of this photoferroionic coupling on charge transport. The asymmetric Schottky barriers at the PtSi/CIPS/Pt interfaces, combined with polarization-dependent barrier modulation, produce a pronounced ferroionic hysteresis and a sweep-rate–dependent current cross-over that reflects slow Cu$^+$ migration. Under illumination, the reduced interfacial barriers and enhanced screening lead to increased carrier injection and a robust photocurrent, consistent with a light-induced lowering of the effective switching barrier. Collectively, these results establish a unified picture in which electronic photoexcitation and ionic polarization dynamics jointly dictate the nanoscale transport and switching behavior of CIPS. Our findings highlight the intrinsic photoferroionic character of CIPS and demonstrate that illumination can serve as an efficient external stimulus for tailoring FE switching thresholds, interfacial barrier profiles, and domain stability in vdW FEs. These insights provide the fundamental design principles necessary for developing next-generation light-programmable FE memories, optoelectronic switches, tunable memristors, and neuromorphic devices based on layered ferroionic materials.

## Supporting Information

Supporting Information includes additional structural, surface potential, electrical, and FE characterization of CIPS. It presents XRD and Raman analyses confirming crystal quality, KPFM measurements of work function and illumination-dependent surface potential, temporal evolution of polarization-induced potential under light, C-AFM topography, current mapping, and I-V curves under light illumination, detailed optoelectronic characterization of CIPS-based FET devices, and comprehensive PFM studies demonstrating thickness-dependent ferroelectric and piezoelectric behavior.

## Author contributions

R.C.B. conceived and supervised the project, and reviewed the manuscript editing. S.C. carried out the experiments and wrote the manuscript. R.B. grew the single crystal and performed the XRD measurement. R.N. deposited the PtSi electrode. All authors have read the manuscript and agreed with its content.

## Acknowledgements

This research has been funded by the United States Department of Defense (DOD) under grant No. W911NF2120213.

## Conflict of interest

The authors declare no conflict of interest.

## Data availability statement

The data that support the findings of this study are available from the corresponding author upon reasonable request.

# Supporting Information

## Photoferroelectric Coupling and Polarization-Controlled Interfacial Band Modulation in van der Waal Compound CuInP$_2$S$_6$

Subhashree Chatterjee, Rabindra Basnet, Rajeev Nepal, and Ramesh C. Budhani*

*Department of Physics, Morgan State University, Baltimore, MD, 21251, USA*

*ramesh.budhani@morgan.edu

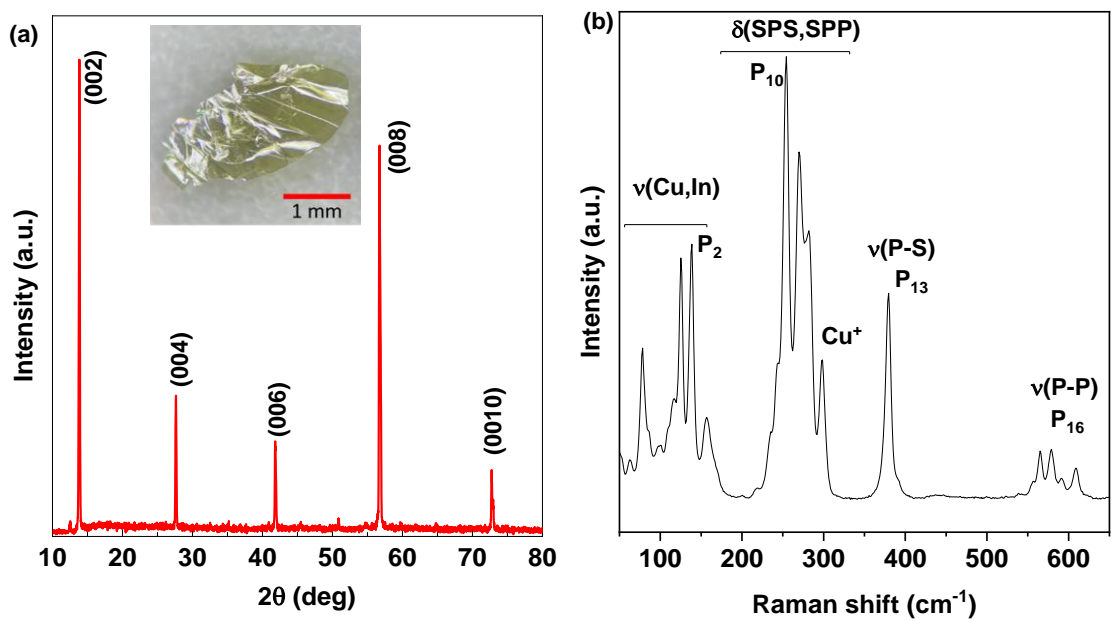

Fig. S1: Room temperature X-ray diffraction pattern, shown in Fig. S1a, confirms the structural quality and orientation of the as-grown crystals. Only (*00L*) reflections appeared, such that the scattering vector was parallel to the c-axis, indicating surface orientation normal to the c-axis. The as-growth, millimeter-sized, green, plate-like single crystals with flat, transparent surfaces are shown in the inset to Fig. S1a. The phase purity of CIPS is also confirmed from the Raman spectroscopy measurement of the as-grown crystal, as shown in Fig. S1b. All the vibrational modes are consistent with the previously reported data.[1]

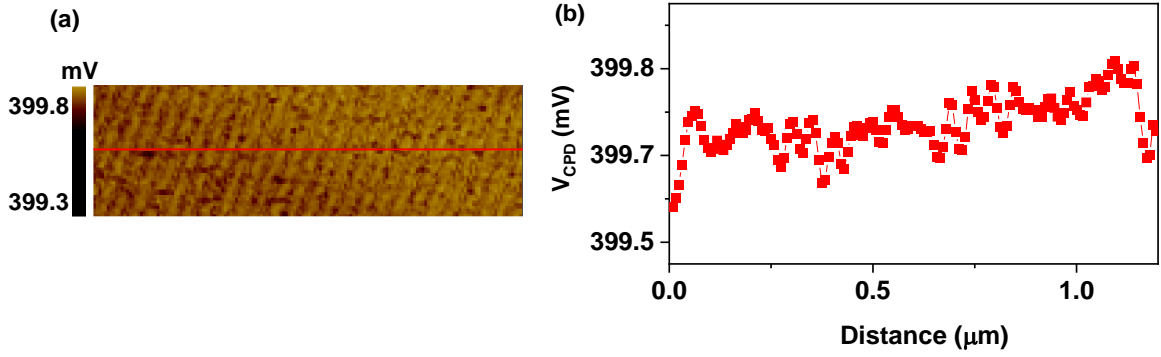

Fig. S2: KPFM $\phi_w$ measurement of Au-coated tip using standard HOPG sample. (a) KPFM surface potential ($V_{CPD}$) of the HOPG sample measured using an Au-coated tip. (b) $V_{CPD}$ line profile measured across the red line shown in (a). $\phi_{w\text{-tip}}$ of the Au tip is measured using the following formula:

$\phi_{w\text{-tip}} = \phi_{w\text{-HOPG}} + eV_{CPD}$

Where, HOPG work function, $\phi_{w\text{-HOPG}} \simeq 4.7$ eV, and $e$ is the electron charge

$V_{CPD} \simeq 0.399$ V

$\phi_{w\text{-tip}} \simeq 5.099$ V $\simeq 5.1$ eV

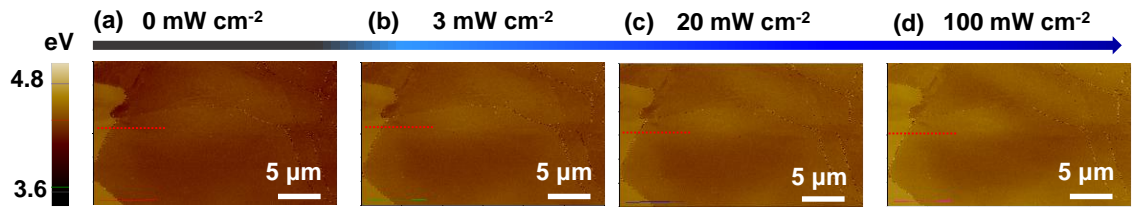

Fig. S3: KPFM $\phi_w$ of CIPS surface measured at different PD varying from 0-100 mW cm$^{-2}$: at (a) dark (0 mW cm$^{-2}$), (b) 3 mW cm$^{-2}$, (c) 20 mW cm$^{-2}$, and (d) 100 mW cm$^{-2}$ illuminations. $\phi_w$ of CIPS is gradually increasing with increasing PD. The $\phi_w$ line profile is measured across the red dotted lines and plotted in Fig. 1d.

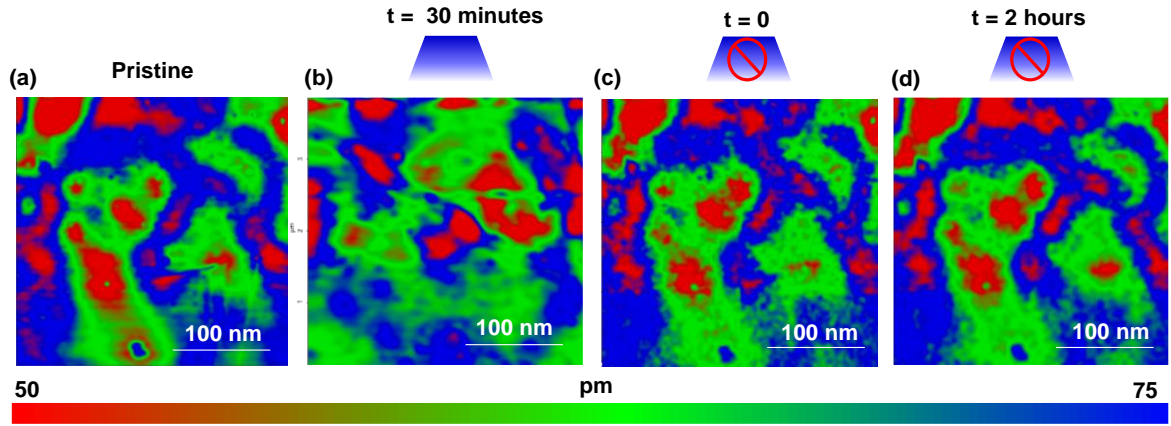

Fig. S4: Temporal evolution of the PFM amplitude images of a selected region of pristine CIPS flake before and after illumination. (d) Pristine CIPS, (e) 30 minutes after continuous light illumination, (f) immediately after the light is switched off, and (g) 2 hours after switching off the light. The PFM amplitude images show relatively smooth spatial variations rather than sharp minima at domain boundaries, likely due to contributions from substrate clamping, tip–sample contact stiffness, and electrostatic interactions influencing the measured electromechanical response.

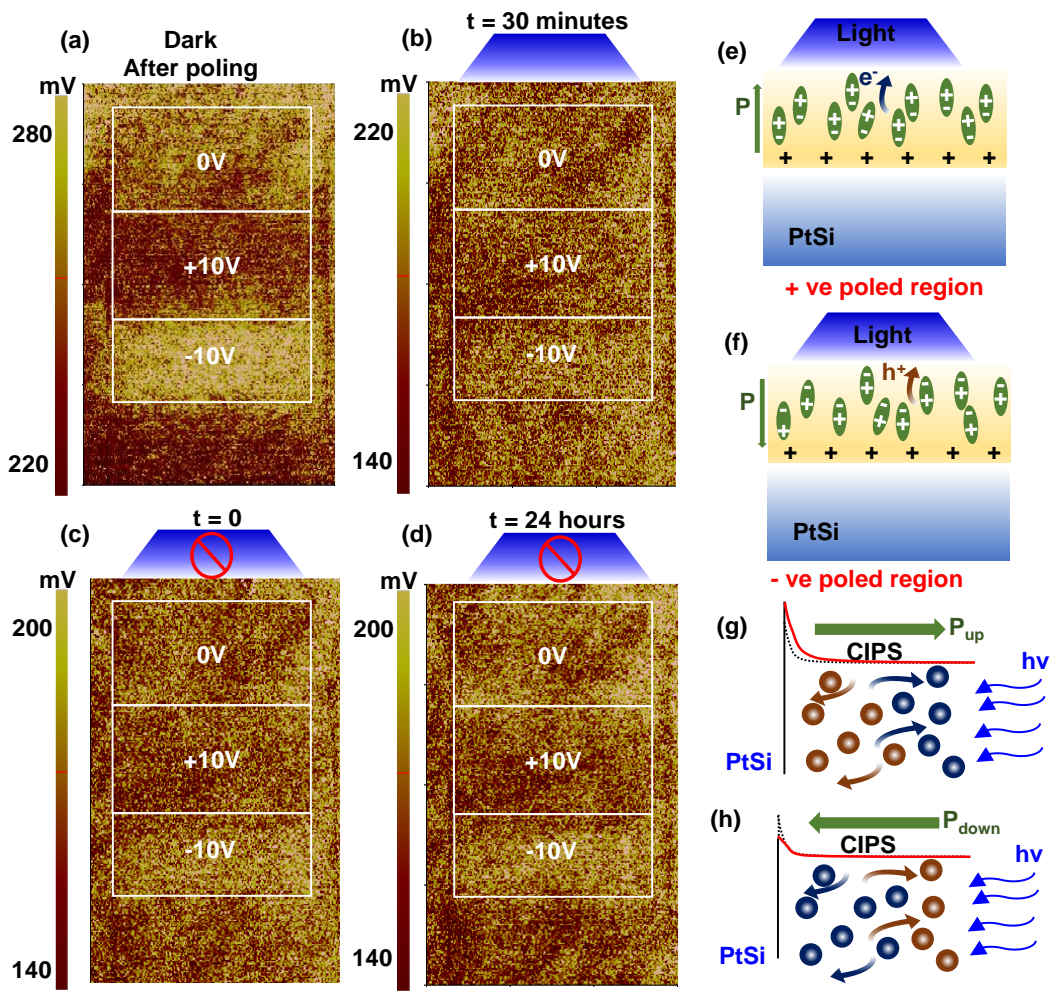

Fig. S5: Temporal evolution of the KPFM $V_{CPD}$ of a 50-nm-thick CIPS flake following local poling with a DC sample bias and subsequent illumination-driven relaxation. (a) $V_{CPD}$ image acquired immediately after poling without illumination, (b) after 30 minutes of continuous illumination in the same area, (c) $V_{CPD}$ measured immediately after the illumination is switched off, and (d) partial restoration of the potential contrast 24 hours after the illumination is removed. Under illumination, the photogenerated surface charge depends on the sign of the surface polarization-bound charge: electrons (e⁻) accumulate in the positively poled regions (e), whereas holes (h⁺) accumulate on the negatively poled regions (f). Here, the brown and blue arrows represent the direction of motion of photo-generated holes and electrons. (g,h) interfacial band bending diagrams and photogenerated carrier migration under illumination (hv) in different pre-poled states. Brown and blue circles represent the photogenerated hole and electrons, respectively.

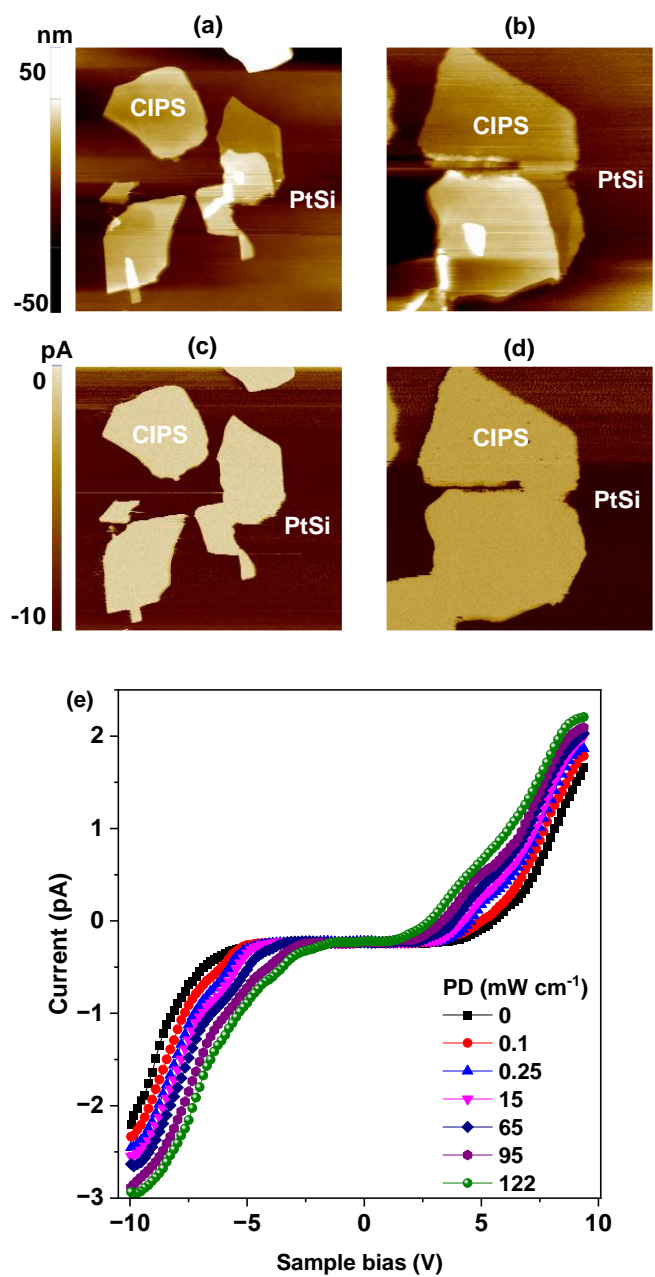

Fig. S6: C-AFM topography and current maps of CIPS layered flakes on a PtSi layer. (a,b) Topographic images of regions 1 and 2. (c,d) Corresponding current maps. (e) Room temperature *I-V* curves measured in C-AFM at different illumination conditions.

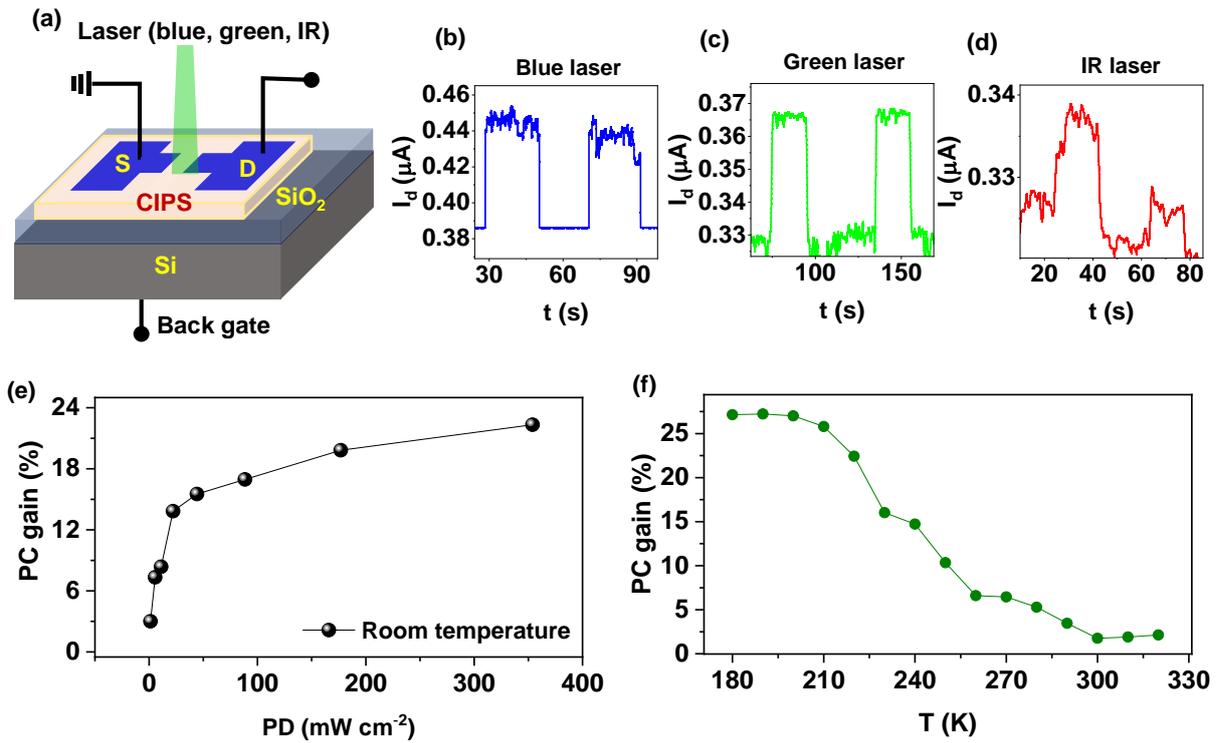

Fig. S7: Device schematic and electrical characterization of a CIPS-based field-effect transistor. (a) Schematic illustration of the CIPS channel device fabricated on a Si/SiO₂ substrate, showing the source (S) and drain (D) electrodes and back-gate configuration. (b–d) Representative drain current responses under light-on/off conditions, under blue, green, and infrared (IR) laser illumination. Photocurrent gain is maximum for the blue laser (445 nm). (e) Room temperature photocurrent gain at different illumination PD. (f) Temperature-dependent photocurrent gain from 180 K to 320 K. The gain increases with decreasing temperature.

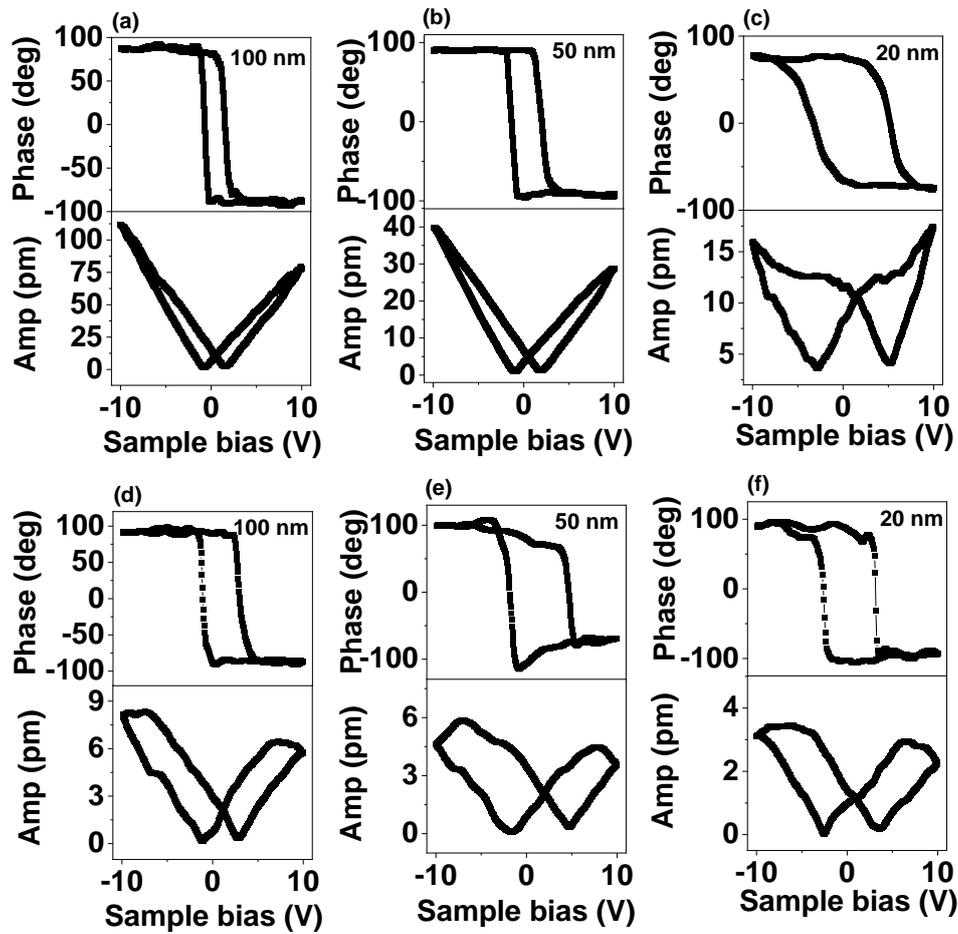

Fig. S8: Room-temperature FE and piezoelectric characterization of a layered CIPS flake by PFM. (a–c) Vertical PFM (VPFM) phase (upper panels) and amplitude (lower panels) hysteresis loops measured in off-state under ±10 V DC bias for CIPS flakes with thicknesses of 100 nm, 50 nm, and 20 nm, respectively. (d–f) Lateral PFM (LPFM) phase (upper panels) and amplitude (lower panels) hysteresis loops measured in off-state for CIPS layers with thicknesses of 100 nm, 50 nm, and 20 nm. The strong response observed in the VPFM measurements indicates a pronounced out-of-plane polarization in CIPS.

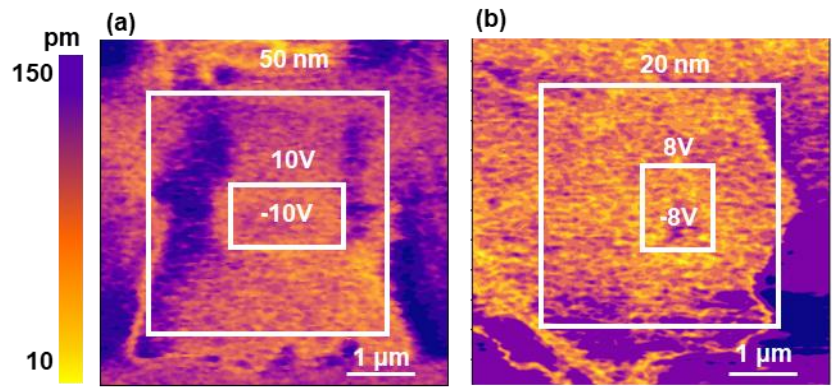

Fig. S9: (a, b) Vertical PFM (VPFM) amplitude image of a CIPS flake with 50 nm, and 20 nm thickness, respectively, after reverse DC bias poling.

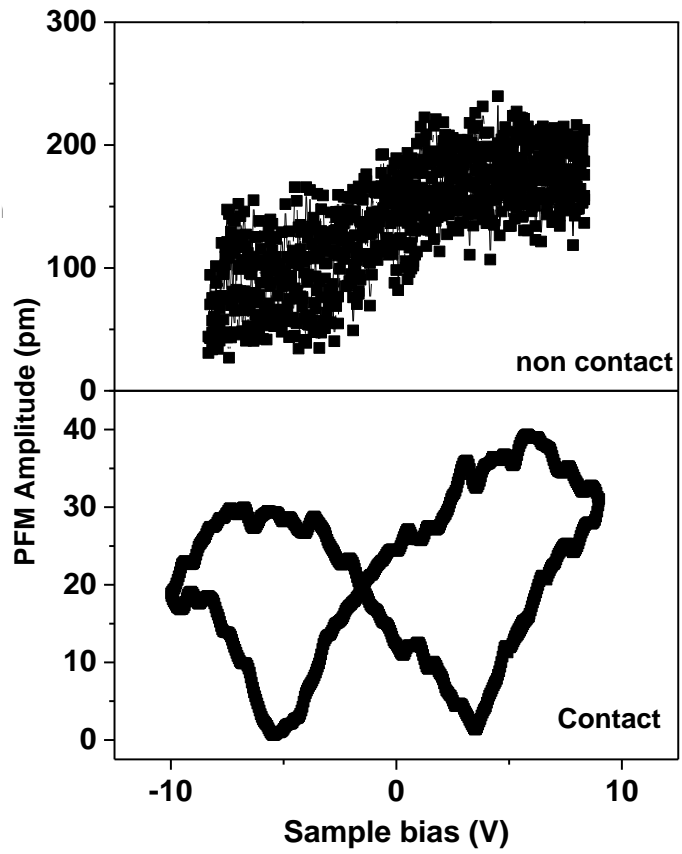

Fig. S10: PFM amplitude measured in a ~50 nm CIPS flake for non-contact (upper panel) and contact (lower panel) mode,[2] revealing true FE behavior in CIPS.